\documentclass[10pt,prl,twocolumn,showpacs,floatfix]{revtex4}

\usepackage{hyperref,amsmath,amssymb,graphicx}
\usepackage[utf8]{inputenc}
\usepackage{t1enc}

\renewcommand{\vec}[1]{\boldsymbol{#1}}
\renewcommand{\d}{\mathrm{d}}

\newcommand{\T}{\mathcal{T}}
\newcommand{\N}{\mathcal{N}}

\newcommand{\bU}{\bar{U}}
\newcommand{\bmu}{\bar{\mu}}

\newcommand{\vphi}{\ensuremath{\varphi}}

\newcommand{\nO}{\ensuremath{ {n_{\text{s}}^{(0)}}}}
\DeclareMathOperator{\La}{L}
\DeclareMathOperator{\Ai}{Ai}

\begin{document}

\title{Quantum Capillary Waves at the Superfluid--Mott Insulator Interface}
\author{Steffen Patrick Rath,$^1$ Boris Spivak,$^2$ and Wilhelm Zwerger$^1$}
\affiliation{$^1$Technische Universit{\"a}t M{\"u}nchen, Physik Department,
James-Franck-Stra{\ss}e, 85748 Garching, Germany\\
$^2$Physics Department, University of Washington, Seattle, Washington 98195, USA
}

\begin{abstract}
We discuss quantum fluctuations of the interface between a superfluid and a
Mott-insulating state of ultracold atoms in a trap. The fluctuations of the
boundary are due to a new type of surface modes, whose spectrum is similar---but
not identical---to classical capillary waves. The corresponding quantum
capillary length sets the scale for the penetration of the superfluid into the
Mott-insulating regime by the proximity effect and may be on the order of
several lattice spacings. It determines the typical magnitude of the interface
width due to quantum fluctuations, which may be inferred from single site
imaging of ultracold atoms in an optical lattice.    
\end{abstract}
\maketitle

The study of fluctuating interfaces is one of the central topics in statistical
physics~\cite{safr2003}, with implications for phenomena like wetting or
capillary forces~\cite{genn2010} or the physics of biological
membranes~\cite{lipo1995}. The origin of interface fluctuations in these cases
is purely thermal.  In recent years, a lot of interest has focussed on phase
transitions that are driven by quantum rather than thermal
fluctuations~\cite{sach1999}.  Somewhat surprisingly, the issue of interface
fluctuations at the boundary between ground states with different order has not
received much attention so far.  In our present work, we study  the interface
between a superfluid (SF) and a Mott-insulator (MI) realized for ultracold
bosons in an optical lattice~\cite{grei2002} as an elementary example of a
quantum interface problem.  Since the MI-SF transition is of second order, the 
coexistence of ground states with different
order in this case is due to the presence of a trapping potential, which gives
rise to a wedding cake structure of successive superfluid and Mott-insulating
domains~\cite{bloc2008}.  This nontrivial spatial structure has been observed in
a direct manner recently by a quantum gas microscope~\cite{bakr2010, sher2010},
which provides single site resolution of individual atoms in an optical lattice.
Within the standard local density approximation (LDA), the SF-MI interface is
sharp. Specifically, for a 2D gas in an isotropic trap, it is a perfect circular
line. Its position is determined by the condition that the local value of the
chemical potential is equal to the critical value for the generic, density
driven SF-MI transition of the homogeneous system~\cite{fish1989} (see
Fig.~1).  As we will show below,  a calculation which incorporates 
fluctuations around a spatially varying smooth background profile of the
superfluid order parameter near the SF-MI boundary gives rise to fluctuations of
this interface. They lead to a quantum uncertainty in its position which can be
described in terms of an effective capillary length $\lambda_g$. The spectrum
$\omega(k)$ of the elementary excitations, which are localized near the
interface, crosses over from a gravity wave like form $\omega(k)=\sqrt{g_{\rm
eff}k}$ at small wave numbers $k\lambda_g\ll 1$ to a free particle like
dispersion $\omega(k)\sim k^2$ at $k\lambda_g\gg 1$.  This is reminiscent of
classical capillary waves, where the role of gravity is played by the external
trap potential. The $k^2$ behavior at short wavelengths is due to the fact that
the SF order parameter vanishes exponentially as one moves into the MI region.
Interactions between the mobile particles thus become negligible. The resulting
free particle dispersion is quite different from the $k^{3/2}$ behavior found
for standard capillary waves, which is due to a non-zero surface tension.  In
the SF-MI case, the latter is zero, however, because the transition is
continuous~\cite{fish1989,sach1999}.  An important feature of this spectrum is
that the amplitude of quantum fluctuations of the position of the SF-MI boundary
diverges as the density gradient goes to zero.  On the qualitative level
mentioned above, the waves are similar to crystallization waves at the rough
superfluid-crystal boundary of $^4$He~\cite{andr1978}.  However, there are
important differences between these cases. The superfluid-crystal transition in
$^4$He is of first order, with
a jump of the density at the boundary.  As a result the crystallization wave
spectrum has a form $\omega\sim k^{3/2}$, and quantum fluctuations of the
boundary position do not diverge. Moreover, recent experiments indicate that the
superfluid-crystal boundary of $^4$He is quantum smooth and that crystallization
waves occur only at finite temperature~\cite{bali2005}.

\begin{figure}[htbp]
  \centering
  \includegraphics[width=0.9\linewidth]{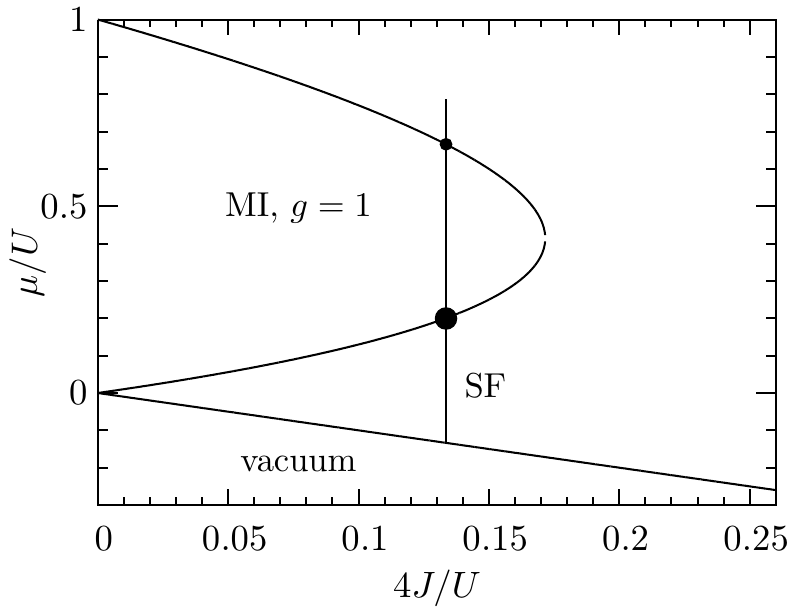}
  \caption{Qualitative zero temperature phase diagram of the homogeneous
  Bose-Hubbard model. In a trap, the local value of the chemical potential
  varies from a maximum in the center to the value where the density vanishes
  (vertical line).  A SF-MI interface appears in the vicinity of a generic
  transition point (marked with black dots). Specifically, we will discuss
  the transition on the lower side of the Mott lobe marked by the heavy dot.}
  \label{fig:phases}
\end{figure}

For a quantitative description of interface fluctuations, we use an effective
action approach, with $\phi(\vec{x},\tau)$ the complex scalar order parameter of
the superfluid~\cite{fish1989}.  At the mean field level, the SF-MI  transition
appears when the dimensionless coefficient $r$ of the quadratic contribution
$r\, |\phi|^2$ to the effective Lagrange density vanishes. In the presence of the trap, $r(\vec{x})$ is spatially
dependent, vanishing at a sharp boundary within LDA. Choosing a coordinate
system where this boundary coincides with the $y$-axis, we have
$r=bx+\ldots $ locally. The coefficient $b$ is determined by the associated gradient
of the chemical potential $\mu$.  Since all other coefficients of
the effective action $S[\phi ]$ for the order parameter are finite near this
boundary,  the relevant model to describe interface fluctuations near the SF-MI
transition is given by
\begin{multline}
  S[\phi]=\int\d^2x\,\d\tilde{\tau}\left\{
\xi_0^2|\nabla\phi(\vec{x},\tau)|^2+bx|\phi(\vec{x},\tau)|^2
\right. \\ \left.
+u|\phi(\vec{x},\tau)|^4
+d\phi^*(\vec{x},\tau)\partial_{\tilde{\tau}}\phi(\vec{x},\tau)\right\}\, .
\label{eq:SLGW}
\end{multline}
Quite generally, the coefficients $\xi_0, b, u$ and $d$ are phenomenological
parameters.  Within a Bose-Hubbard model description of the SF-MI transition,
they can be calculated directly from the hopping and interaction energy
parameters $J$ and $U$ of the microscopic Hamiltonian, using the dimensionless
function~\cite{vano2001,seng2005} 
\begin{equation}
  \chi_0(\bmu,\bU)=\frac{g}{\bmu-(g-1)\bU}+\frac{g+1}{g\bU-\bmu}\, .
  \label{eq:chi0}
\end{equation}
Here, $g=1,2,\ldots$ is the integer density corresponding to a specific Mott
lobe, while $\bmu=\mu/4J$ and $\bU=U/4J$. One generally has 
$b=-4\partial_\mu\chi_0\partial_x\mu$ and $d=-\partial_{\bmu}\chi_0$. The
general expression for $u$ is quite unwieldy, but reduces to $u\approx 4l^2$
in the regime $U\gg J$. Note that for the
lower boundary of the one-atom Mott lobe, (see
Fig.~1) $\partial_{\bmu}\chi_0<0$ . Moreover,
$\tilde{\tau}=J\tau$ is the dimensionless time, while the scale for $u$ and 
$\xi_0=l$ is simply the lattice spacing. For $J/U$ close to the
critical value near the tip of the lobe, $b$ and $d$ vanish so that higher order
terms in $x$ must be taken into account for $b$ while the $d$ term must be
supplemented with the second time derivative term following from the gradient
expansion of the action. Quite generally,
the characteristic length $\xi_0$ defines a bare length scale such that the bulk
correlation length is $\xi=\xi_0/\sqrt{r}$ in a homogeneous system, diverging as
$(\mu-\mu_c)^{-1/2}$ at the transition. Note that we are considering the problem
in two dimensions, which is the upper critical dimension since the dynamical
exponent is $z=2$ for the generic SF-MI transition~\cite{fish1989, sach1999}.
As a result, mean field theory correctly describes the divergence of the
correlation length up to logarithmic corrections. In the presence of an external
trap potential, the coefficient $r$ vanishes linearly in the vicinity of the
sharp SF-MI interface that results within LDA.  The characteristic length scale
$\lambda_g$ over which this sharp profile will be smeared out by fluctuations is
determined by the condition $\xi_0/\sqrt{r(\lambda_g)}=\lambda_g$. It identifies
$\lambda_g$ with the  scale at which the local correlation length reaches
$\lambda_g$ itself~\cite{ginz1977}. This results in a broadening
of the interface over a scale $\lambda_g=(\xi_0^2/b)^{1/3}$, a result that is
borne out in detail by our calculation below.

\begin{figure}[htbp]
  \centering
    \includegraphics[width=.9\linewidth]{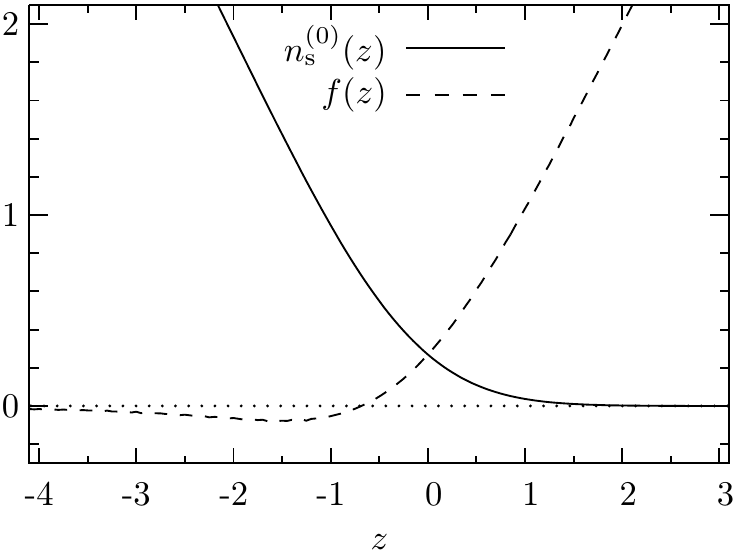}
  \caption{The mean field superfluid density $\nO$ and the function $f$
  appearing in the differential equations for $\T$ and $\N$ as functions of $z$,
  in dimensionless units.
 }
  \label{fig:n0f}
\end{figure}

The equilibrium order parameter profile can be obtained by solving the 
Euler-Lagrange equation corresponding to the action which reads
\begin{equation}
  \tilde{\phi}''-z\tilde{\phi}-|\tilde{\phi}|^2\tilde{\phi}=0\ ,
  \label{eq:mfeq}
\end{equation}
where we have defined $z=x/\lambda_g$ and $\tilde{\phi}=\phi/\phi_g$ with
$\phi_g=\sqrt{b\lambda_g/2u}$. This equation does not have a closed-form
solution, but one readily finds the asymptotic behavior, i.e.,
$\tilde{\phi}\sim\sqrt{-z}$ for $-z\gg 1$. For $z\gg 1$, the nonlinear term
becomes negligible and Eq.~\eqref{eq:mfeq} becomes the Airy differential
equation which is solved by $\tilde{\phi}=\Ai(z)$. For arbitrary $z$, the
solution can be obtained numerically. The resulting order parameter profile
$\nO(z)=|\tilde{\phi}|^2$ is shown in Fig.~\ref{fig:n0f} and is
formally identical to that obtained at a SF-vacuum
boundary~\cite{dalf1996,lund1997,fett1998}. Note however that in the present
case the characteristic length $\lambda_g$ is different. In particular, it
depends explicitly on the interactions through the function $\chi_0$. Unlike the
LDA, the mean field solution predicts a smooth transition between insulating
and superfluid regions over the length $\lambda_g$. Since all coefficients in
Eq.\eqref{eq:mfeq} are real, the mean field solution $\phi_0$ can be chosen real
without loss of generality.

In the following, we consider quantum fluctuations around this mean field order
parameter profile. To this end, we expand the action \eqref{eq:SLGW} to second
order in deviations from the mean field solution $\phi_0$. When $\phi_0$ is
small, it is natural to consider fluctuations of the real and imaginary part of
$\phi$, i.e., $\phi=\phi_0+\vphi+i\psi$. Conversely, where $\phi_0$ takes
appreciable values it is more natural to take into account the $U(1)$ symmetry
of the action and consider fluctuations of $n_{\text{s}}$ and $\theta$, where
$\phi=\sqrt{n_{\text{s}}}e^{i\theta}$. In both cases, fluctuations take on the
form of plane waves parallel to the interface while in the direction
perpendicular to the interface they form a set of modes which must be determined
from a solution of the Euler-Lagrange equations. The equations for the
dimensionless versions of $\varphi$ and $\psi$ (for ease of notation, we omit
the tilde in the following) read
\begin{equation}
  \begin{split}
    \varphi'' - (z+3\nO+k^2)\varphi + i\omega\psi = 0
    \\
    \psi'' - (z+\nO+k^2)\psi-i\omega\varphi = 0\ ,
  \end{split}
  \label{eq:coupledReIm}
\end{equation}
with $\nO=\phi_0^2$.
In the density-phase representation it is convenient to define
$\T=\phi_0\theta$ and $\N=\delta n_{\text{s}}/\phi_0$ which obey the coupled
equations 
\begin{equation} 
  \begin{split} 
    \T''-(f+k^2)\T-i\omega \N = 0 \\
  \N''-(f+2\nO+k^2)\N+i\omega \T = 0\ , 
\end{split} 
\label{eq:couplednt}
\end{equation}
where $f=\partial_z^2\nO/2\nO-(\partial_z\nO/2\nO)^2$. Asymptotically, $f\sim z$
for $z\gg 1$ and $f\sim-1/4z^2$ for $-z\gg 1$. The function $f(z)$ is shown
in Fig.~\ref{fig:n0f}.

In both representations, we have two coupled second order differential equations
with non-constant coefficients. Analytical solutions may thus be expected only
in limiting cases. There are two opposite regimes where the equations can be
solved analytically: firstly, the limit where the wavelength of the interface
fluctuations is small compared to $\lambda_g$, and secondly, the opposite case
where the wavelength is much larger than the size of the fluctuation region.

The first case turns out to correspond to the limit of low superfluid densities
where one can neglect the $\phi^4$ term in the action \eqref{eq:SLGW}.
Fluctuations then have the same general structure as the mean field solution,
i.e.,
\begin{equation}
 \phi(z)=\Ai[z-(\omega-k^2)]e^{i(ky-\omega t)}\ . 
  \label{eq:dilutephi}
\end{equation}
This fixes the frequency $\omega$ up to an additive constant $\omega-k^2$
which must be determined by comparison with a solution to the complete set of
differential equations.

The second case, the long wavelength limit, is just the Thomas-Fermi
approximation which is familiar from the calculation of hydrodynamic modes in
Bose-Einstein condensates \cite{stri1996}. In this limit, one obtains (with
$\theta(y,z)=\hat{\theta}(z)e^{i(ky-\omega t)}$)
\begin{equation}
  z\hat{\theta}''+\hat{\theta}'-(k^2x+\omega^2/2)\hat{\theta}=0
  \label{eq:deqnt}
\end{equation}
and the same equation for the superfluid density fluctuations. This equation has
a complete orthonormal set of solutions
\begin{equation}
  \hat{\theta}_{nk}(z)=\sqrt{2k}e^{kz}\La_n(-2kz)\ ,
  \label{eq:TFmodes}
\end{equation}
where $\La_n$ are the Laguerre polynomials, $n=0,1,\dots$. These modes, which
are defined for $z\leq 0$ only (the Thomas-Fermi approximation to the mean field
profile vanishes identically for $z>0$), correspond to the dispersion law
(substituting all constants to obtain a dimensionful quantity)
\begin{equation}
  \omega_{n,k}=\sqrt{\frac{2\xi_0^2b(2n+1)}{d^2}k}
  \equiv \omega_g\sqrt{2(2n+1)k\lambda_g}\ ,
  \label{eq:TFdisp}
\end{equation}
i.e., the solutions can be grouped into ``branches'' characterized by the
integer $n$, each branch having a gravity wave like $\sqrt{k}$ dispersion. To
make this analogy more explicit, one may write the lowest branch as
$\omega=\sqrt{g_{\text{eff}}k}$, with $g_{\text{eff}}=2\xi_0^2b/d^2$. An
equivalent dispersion relation has been derived in the context of the boundary
of a dilute Bose-Einstein condensate in \cite{alkh1999}.

\begin{figure}[htbp]
  \centering
  \includegraphics[width=\linewidth]{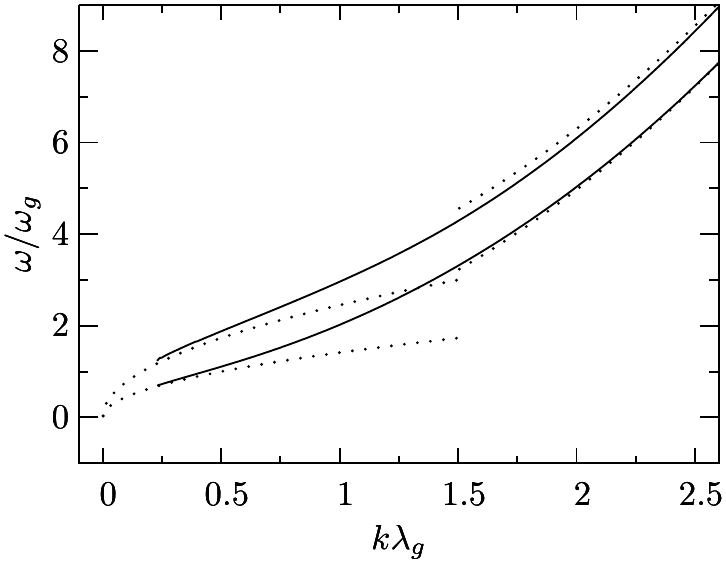}
  \caption{The two lowest branches of the dispersion relation
  $\omega_n(k)$ as obtained from a numerical calculation (solid lines). The
  asymptotic behavior for small and large $k$ is indicated with dotted lines.}
  \label{fig:dispersion}
\end{figure}

From these limiting cases, we see that the dispersion relation $\omega(k)$
starts as $\omega\sim\sqrt{k}$ at small $k$ and then gradually crosses over to a
$\omega\sim k^2+\text{const}$ behavior as $k$ increases. To see how this
crossover happens and to determine the additive constant for each branch in the
large $k$ regime, we have numerically solved the coupled sets of differential
equations \eqref{eq:coupledReIm} and \eqref{eq:couplednt} using a matrix Numerov
method~\cite{gord1969}.  The result is shown in Fig.~\ref{fig:dispersion}.  The
additive constants for the lowest two branches are determined by a least square
fit to the upper part of the dispersion curves as $0.97(2)$ and $2.27(9)$. The
results are in agreement with a prior numerical solution of the same equations
in the context of the boundary of a dilute BEC, where the lowest band dispersion
was derived using a finite difference method~\cite{angl2001}. The crossover in
the lowest band indeed happens at $k\lambda_g\approx 1$ so that $\lambda_g$
plays a role analogous to that of the capillary length in the physics of surface
waves on deep water. Note that this is completely different from the results of
a previous analysis of boundary fluctuations of the SF-MI interface by Mariani
and Stern~\cite{mari2005}, who found a large $k$ scaling $\omega(k)\sim k^{3/2}$
based on a model with a phenomenological nonzero surface tension.

To estimate the spatial extension of the fluctuation region, we introduce the
phenomenological height variable $h(y,t)=\lambda_g\delta
n(y,0,t)/\partial_z\nO$. The mean square fluctuations $\langle h^2\rangle$ would
diverge if there was only the $k^2$ part of the dispersion, but are made finite
due to the crossover to gravity waves $\omega\sim\sqrt{k}$ precisely as in the
case of classical capillary waves. In the long wavelength regime, the action
can be parametrized using the modes \eqref{eq:TFmodes} and then reads
\begin{equation}
  S=\frac{\beta \xi_0^2d}{2u}\sum_{k,m,n}
  \begin{pmatrix}\theta_{k,m,-n} \\ \delta n_{k,m,-n}\end{pmatrix}
    \begin{pmatrix}
      \omega_{m,k}^2/2 & -\omega_n \\
      \omega_n & 2
    \end{pmatrix}
    \begin{pmatrix}
      \theta_{k,m,n}\\ \delta n_{k,m,n}
    \end{pmatrix}\ ,
  \label{eq:paramactionTF}
\end{equation}
where $\omega_n=2\pi n/\beta\omega_g$ are the bosonic Matsubara frequencies,
rendered dimensionless. Note that the
combination $\xi_0^2d/u$ of the dimensionful Landau-Ginzburg coefficients is
dimensionless. This representation allows to calculate the variances of 
$\theta$ and $\delta n$. A lower bound to the total fluctuations is then given
by the contribution from the lowest branch
and wavelengths smaller than $\lambda_g$: 
\begin{equation}
  \langle h^2\rangle \gtrsim
  \langle h^2\rangle_< = \frac{2\sqrt{2}}{5\pi}\frac{u}{\xi_0^2d}\lambda_g^2
  \label{eq:smallfluct}
\end{equation}
i.e., the fluctuations are infrared convergent so that the fluctuation region
remains confined to the mean field transition region and the interface is
quantum smooth.

To estimate whether the predicted zero point fluctuations of the SF-MI interface
may be observed with current experiments, we
take the typical case of $^{87}\text{Rb}$ atoms in an optical lattice of period
$l=532\,\text{nm}$, a lattice height of $V_0=16.4E_r$ which is close to the
transition point for the MI-SF transition at unit filling $g=1$
($E_r=h^2/2m\lambda_{\text{lat}}$ is the recoil energy) and a central
chemical potential $\mu(0)=1.1\mu_c$. With these parameters, the effective
capillary length $\lambda_g$ is equal to one lattice spacing for an isotropic
harmonic trap with frequency $2\pi\times 16.3\,\text{Hz}$. Moreover, the
characteristic frequency $\omega_g=\sqrt{g_{\text{eff}}/\lambda_g}$ where the
spectrum crosses over from a gravity-like form $\sim\sqrt{k}$ to the free
particle $\sim k^2$ regime is about $2\pi\times 9\,\text{Hz}$. To see quantum 
fluctuations of the interface requires the temperature to be smaller than
$\hbar\omega_g/k_{\text{B}}=0.43\,\text{nK}$. This is quite challenging to reach
but appears feasible with novel cooling techniques like spin gradient
demagnetization, where temperatures around $0.35\,\text{nK}$ have recently been
achieved in a similar setup~\cite{medl2011}. Note that the regime
$\hbar\omega_g>k_{\text{B}}T$ that is required to see quantum fluctuations of
the interface is opposite to the standard semiclassical limit
$\hbar\omega\ll k_{\text{B}}T$ that is usually considered in the thermodynamics
of trapped BECs~\cite{gior1997}. Moreover, for the
parameters above,  the characteristic density at the transition is
$\phi_g^2l^2\approx 0.125$ so that [multiplying by $\nO(0)$] the density of
mobile holes in the transition region is about $0.034$ per lattice site.  Using
a quantum gas microscope~\cite{bakr2010,sher2010}, the smooth non-LDA mean field
profile can be measured by averaging over a sufficiently large number of images.
To probe the dispersion relation, one needs to selectively excite individual
modes which can equally be achieved thanks to the single-site addressability of
quantum gas microscopes: by modulating, e.g., the lattice depth on the
single-site level, one can achieve values of $2\pi/k$ ranging from $4\pi R$
(corresponding to the quadrupole mode, with $R$ the radius of the LDA transition
circle) down to $2l$.  Since $\lambda_g$ can take any value from practically
zero to several lattice sites, this permits to map out the crossover shown in
Fig.~\ref{fig:dispersion}. With the stated parameters, the experimental
detection of the interface's quantum dynamics is certainly challenging, but
within reach of current experimental technology. Additional flexibility may be
gained from the use of non-harmonic potentials using phase plates. For example,
a box-like potential with an added localized strong variation permits to have
the transition happen at a larger radius so that more atoms participate.

In summary, we have discussed the zero temperature quantum fluctuations of the
MI-SF interface and found that the  associated dispersion relation
leads to a quantum smooth surface. An experimental observation of these
fluctuations requires very shallow trap potentials, where the associated
capillary length $\lambda_g\sim\omega^{-1/3}$ is at least several lattice
spacings. From a more
general point of view, the interface fluctuations we discussed here are just a
particular case of the rich physics of interfaces in quantum phase transitions.
For example, in recent years, a completely new type of interfaces has turned
into the focus of research, in which novel phases appear at the boundary
between two materials with different ground states. A striking example is the
appearance of a conducting 2D electron gas and even tunable superconductivity at
the boundary between two insulators~\cite{reyr2007,cavi2008}. 

We are grateful for helpful discussions with I.~Bloch, F.~Gerbier and
M.~Greiner.  B.~S.\ acknowledges financial support from the Humboldt foundation,
S.~P.~R.\ and W.~Z.\ from the DFG within the Forschergruppe 801.

\end{document}